\documentclass{desyproc}

\begin{document}
\title{Recent results of search for solar axions using resonant absorption by $^{83}$Kr nuclei}

\author{{\slshape A.V. Derbin$^{2}$, I.S. Drachnev$^{2}$, A.M. Gangapshev$^{1}$, Yu.M. Gavrilyuk$^{1}$, V.V. Kazalov$^{1}$, V.V. Kobychev$^{4}$, V.V. Kuzminov$^{1}$,  V.N. Muratova$^{2}$, S.I. Panasenko$^{3}$,
S.S. Ratkevich$^{3}$, D.A. Tekueva$^{1}$, E.V. Unzhakov$^2$, S.P. Yakimenko$^{1}$} \\ [1ex]
$^1$ Institute for Nuclear Research, RAS, Moscow, Russia \\
$^2$ NRC "Kurchatov Institute" Petersburg Nuclear Physics Institute, St. Petersburg, Russia \\
$^3$ Kharkov National University, Kharkov, Ukraine \\
$^4$  Institute for Nuclear Research of NAS Ukraine, Kiev, Ukraine \\}

% if the proceedings are available online (e.g. at Indico)
% please enter the contribution ID or file_name below for the DOI
%\contribID{32}

\contribID{derbin\_alexander}

	% TO THE CONFERENCE EDITORS: 
% please update the following information      
% before sending the template to the authors

\confID{13889}  % if the conference is on Indico uncomment this line
\desyproc{DESY-PROC-2017-XX}
\acronym{Patras 2017} % if you want the Acronym in the page footer uncomment this line
\doi  % if there is an online version we will register DOIs
\maketitle

\begin{abstract}
A search for resonant absorption of the solar axion by $^{83}\rm{Kr}$ nuclei was performed using the proportional counter
installed inside the low-background setup at the Baksan Neutrino Observatory.  The obtained model independent upper limit on the
combination of isoscalar and isovector axion-nucleon couplings $|g_3-g_0|\leq 8.4\times 10^{-7}$  allowed us to set the new
upper limit on the hadronic axion mass of $m_{A}\leq 65$ eV (95\% C.L.) with the generally accepted values $S$=0.5 and $z$=0.56.
\end{abstract}

\section{Introduction}

%If axions do exist, then the Sun should be an intense source of these particles. In 1991 Haxton and Lee calculated the energy
%loss of stars along the red-giant and horizontal branches due to the axion emission in nuclear magnetic transitions in
%$^{57}\rm{Fe}$, $^{55}\rm{Mn}$, and $^{23}\rm{Na}$ nuclei \cite{HaxLee}. In 1995 Moriyama  proposed experimental scheme to search
%for 14.4 keV monochromatic solar axions that would be produced when thermally excited $^{57}\rm{Fe}$ nuclei in the Sun relax to
%its ground state and could be detected via resonant excitation of the same nuclide in a laboratory \cite{Mor95}. Searches for
%resonant absorption of solar axions emitted in the nuclear magnetic transitions were performed with $^{57}\rm{Fe}$, %$^{7}\rm{Li}$,
% $^{169}\rm{Tm}$ and $^{83}\rm{Kr}$ (see \cite{Der11} and refs therein).

%In this paper we present new results of the search for solar axions using the resonant absorption by $^{83}\rm{Kr}$ nuclei
%\cite{Gav15}. The energy of the first excited $7/2^+$ nuclear level  is equal to 9.405 keV, lifetime $\tau = 2.23\times10^{-7}$
%s, internal conversion coefficient $\alpha = 17.0$ and the mixing ratio of $Ì1$ and $Å2$ transitions is $\delta$ = 0.013.

If the axion exists, the Sun should be one of the most intense sources of  these particles. 
The aim of this work is to search for monochromatic axions with an energy of 9.4 keV emitted in the M1 transition in the $^{83}\rm{Kr}$ nuclei in the Sun \cite{Gav15}. 
Axions on the Earth can be detected in the inverse reaction of resonance absorption by detecting particles ($\gamma$- and X-ray photons, as well as conversion and Auger electrons) appearing at the decay of an excited nuclear level. 
The probability of the emission and subsequent absorption of axions depends only on the coupling constant with nucleons, which is minimally model dependent and is proportional to axion-nucleon coupling constant $(g_{AN})^4$.

The axion flux was calculated in \cite{Gav15} for the standard solar model BS05 \cite{Bah05} characterized by a highmetallicity \cite{Gre98}. 
The differential flux at the maximum of the distribution is \cite{Gav15}:
\begin{equation}\label{axionflux_num}
\Phi_{A}(E_{M1}) = 5.97\times 10^{23}\left(\frac{\omega_{A}}{\omega_{\gamma}}\right) \rm{cm}^{-2} \rm{s}^{-1}
\rm{keV}^{-1}.
\end{equation}

where ${\omega_{A}}/{\omega_{\gamma}}$ is the branching ratio of axions to photons emission.
The cross section for resonance axion absorption is given by an expression similar to the expression
for the photon-absorption cross section, the correction for the ratio $\omega_A /\omega_{\gamma}$ being taken into
account.

\begin{equation}\label{crosssection}
\sigma(E_{A})=2\sqrt{\pi}\sigma_{0\gamma}\exp\left[-\frac{4(E_{A}-E_{M})^{2}}{\Gamma^{2}}\right]\left(\frac{\omega_{A}}{\omega_{\gamma}}\right),
\end{equation}

where $\sigma_{0\gamma}$ is the maximum cross section of the $\gamma$ -ray resonant absorption and
$ \Gamma= 1/\tau$. 
The total cross section for axion absorption can be obtained by integrating $\sigma(E_A)$ over the axion
spectrum. 
The expected rate of resonance axion absorption by the $^{83}\rm{Kr}$ nucleus as a function of the ratio
$\omega_A/\omega_{\gamma}$, the combination of isoscalar and isovector coupling constants $|g_{3} - g_{0}|$ and axion mass $m_A$ can be represented in the form ($S$ = 0.5, $z$ = 0.56)\cite{Gav15}:
\begin{eqnarray}\label{count_speed}
R_A \rm{[g^{-1}day^{-1}]} = 4.23\times10^{21}(\omega_{A}/\omega_{\gamma})^2 \\ \label{count_speed_2} =
8.53\times10^{21}(g_3-g_0)^4(p_A/p_{\gamma})^6
\\ \label{count_speed_3}  = 2.41\times10^{-10}(m_{A}/\rm{1~eV})^{4}(p_A/p_{\gamma})^6.
\end{eqnarray}

%\begin{wrapfigure} {l}
%\centerline{\includegraphics[bb = 100 450 520 755, width=0.45\textwidth, %height=0.3\textheight]{Derbin_Alexander_fig1.pdf}}
%\caption {Original energy spectrum and spectrum after rejection of the events with pulse rise  time $\geq 3.8 \mu$s and %$\lambda
%\leq 0.115$} \label{Derbin_Alexander_fig1.pdf}
%\end{wrapfigure}

\begin{figure}[ht!]
	\centerline{\includegraphics[width=0.45\textwidth, height=0.35\textheight]{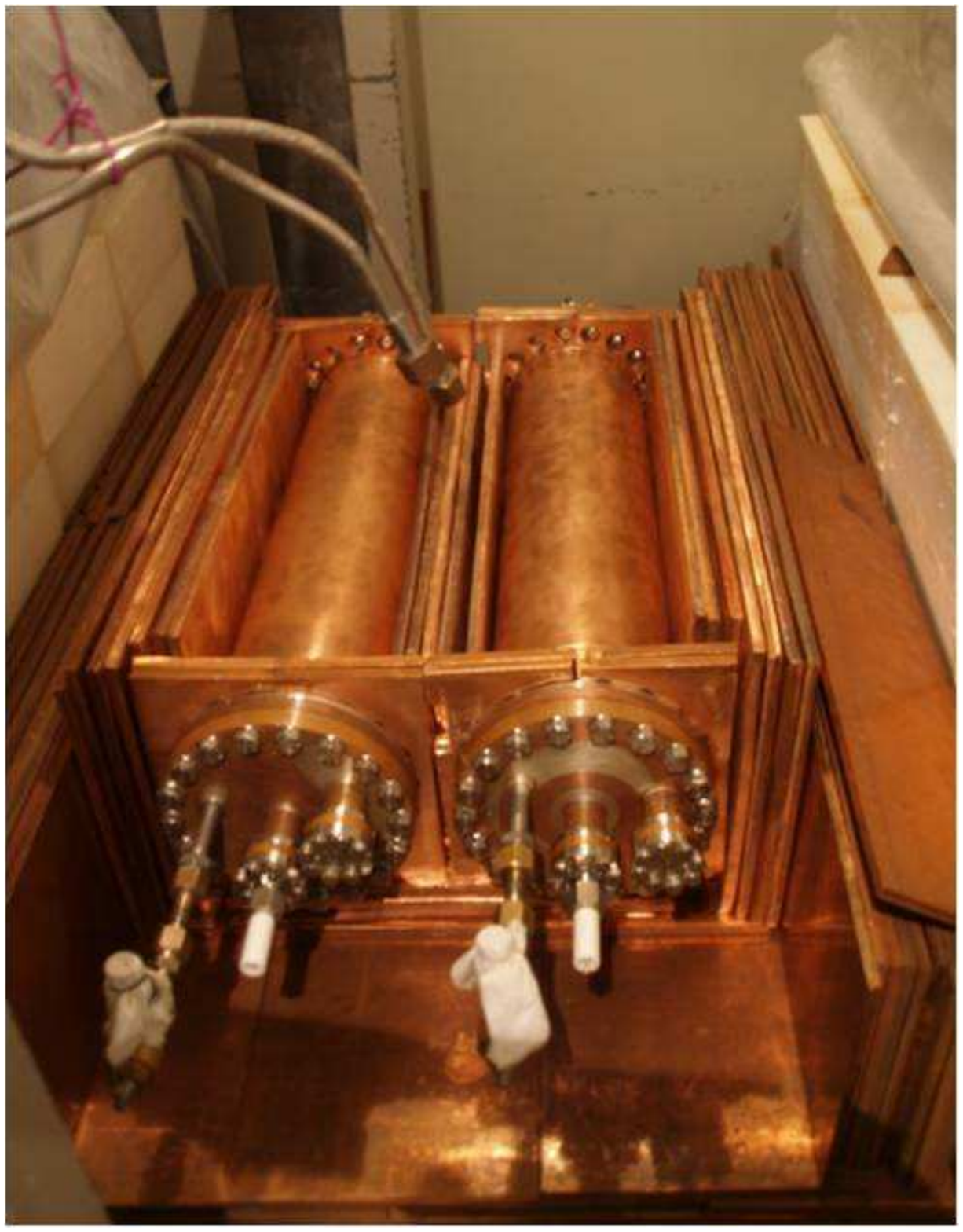}}
	\caption{A large proportional counter (LPC) with a casing of copper inside passive shilding. }
	\label{fig:1}
\end{figure}

\section{Experimental setup}

A large proportional counter filled with of $^{83}\rm{Kr}$ (99.9$~\%$)  was used to detect X-rays and gammas, as well as conversion and Auger electrons, appearing in the decay of the excited level with an energy of 9.4 keV. The LPC is a cylinder with inner diameters of $137$ mm. 
A gold-plated tungsten wire of 10 $\mu$m in diameter is stretched along the LPC axis and is used as an anode. 
In order to reduce the influence of edge effects on the collection of the charge, the ends of the anode wire were surrounded by copper tubes (3 mm in diameter and 38 mm in length), which were at the anode potential and excluded gas amplification in this region. 
With the inclusion of Teflon insulators, the distance from the working area to the flanges of the chamber was 70 mm. The length of the working area of the chamber was 595 mm, which corresponded to a volume of 8.77 L. 
The chamber operated at a pressure of 1.8 bar. The mass of $^{83}\rm{Kr}$ isotope in the working volume was 58 g.

 The LPC is surrounded by passive shield made of copper ($\sim$20 cm), lead ($\sim$20 cm) and  polyethylene (8 cm).  
 The setup is located in the Deep Underground Low-Background Laboratory of Institute for Nuclear Research of Russian Academy of Sciences (BNO INR RAS)\cite{DULB}, at the depth of 4900 m w.e., where the cosmic muon flux is reduced by $\sim 10^7$ times in comparison to that above ground, and evaluated as $(2.6 \pm 0.09) \times 10^{-9}$~cm$^{-2}$s$^{-1}$.% \cite{Gav}.

A signal from the anode was supplied to a charge-sensitive preamplifier. 
The shape of the pulse was digitized in a time interval of 164 $\mu s$ with a frequency of 12.5 MHz and was transmitted to a computer through a USB port. 
The rise time of the leading edge of the pulse and the ratio of amplitudes of secondary and primary pulses were determined for each event, because these parameter makes it possible to select events near the cathode and nonpoint events such as multiple Compton scattering. 
The procedure of the analysis of the shape of pulses was described in more detail in \cite{PTE}.

\begin{figure}[ht!]
	\centerline{\includegraphics[width=0.45\textwidth, height=0.3\textheight]{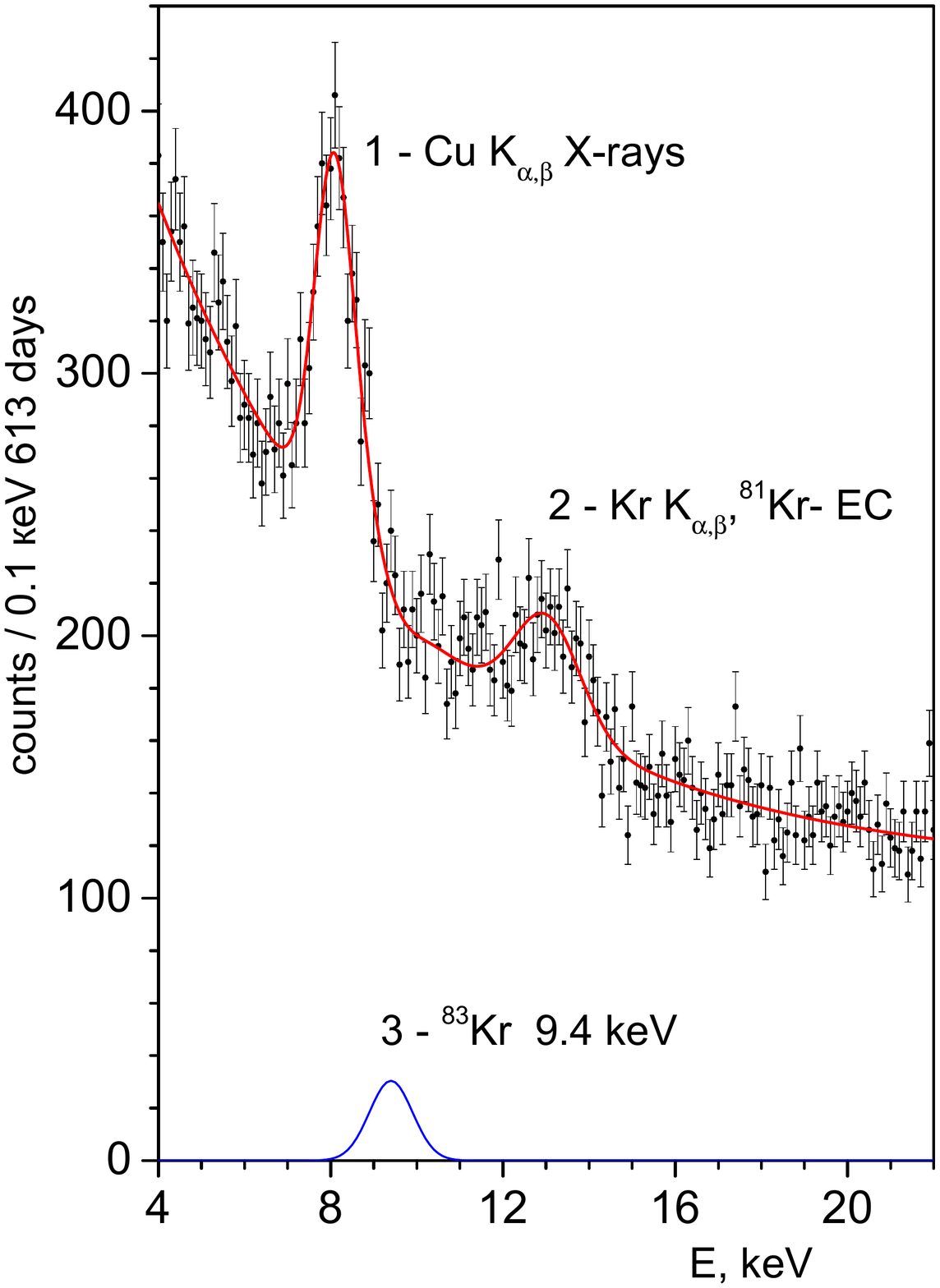}}
	\caption{Energy spectra of the Kr LPC measured for 613 days, fitting results (red line) and expected axion peak for 3$S_{lim}$ (blue line). }
	\label{fig:2}
\end{figure}

\section{Results}

The background spectra collected during 613.25 days and fit result curve are presented in Fig.\ref{fig:2}. 
Two peaks are clear visible in the energy range (4-26) keV. 
The peak with energy 8.05 keV associates with the detection of $\rm{K}_{\alpha1,2}$ X-rays of copper. The structure of the second peak is more complicated, it is mixture of Kr and Br $\rm{K}_{\alpha1,2}$ X-rays and 13.5 keV from $K$-capture of cosmogenic $^{81}$Kr.
%The distributions of the events versus pulse rise time and the ratio of amplitudes of secondary and primary pulses were investigated %\cite{PTE}. 
It is seen that the 9.4 keV peak is not manifested. 
The maximum likelihood method was used to determine the intensity of the peak.
The fit of spectrum  corresponding to the minimum  $\chi^2$ is shown by red solid line in Fig. 2.
The minimum of $\chi^2$ corresponds to the nonphysical value of the area of the 9.4 keV peak $S_A = -(102 \pm 92)$
events.
The standard  $\chi^2$-profile method was used to determine the upper bound on the number of events in the peak. 
The upper bound thus determined for the number of events in the peak is $S_{lim}$ = 127 for 95$~\%$ C.L.

The expected number of registered axions is 
\begin{equation}\label{CRate}
S_A = RMT\epsilon \leq S_{lim},
\end{equation}
where $M$ = 58 g is mass of $^{83}\rm{Kr}$ isotope, $T$ = 613.25 days is time of data taking, and  $\epsilon = 0.825$ is  the detection efficiency. 

The upper limit on the excitation rate of $^{83}$Kr by solar hadronic axions is defined as $R_{exp}=4.29\times 10^{-3}~~\rm{g^{-1}day^{-1}}$.    The relation $R_A \leq R_{exp}$ limits the region of possible values of the coupling constants $g_0$, $g_3$ and axion mass $m_A$. 
In accordance with Eqs. (\ref{count_speed}-\ref{count_speed_3}), and on condition that $(p_A/p_\gamma)\cong 1$ provided for $m_A < 3$ keV one can obtain:
\begin{equation}\label{limwawg}
(\omega_{A}/\omega_{\gamma}) \leq 1.0\times 10^{-12},
\end{equation}
\begin{equation}\label{limgAN}
|g_3-g_0|\leq 8.4\times 10^{-7}, \rm{~and}
\end{equation}
\begin{equation}\label{limma}
m_A \leq 65 \rm{~eV~~ at~ 95\%~ C.L.}
\end{equation}

The limit (\ref{limma}) is stronger than the constrain obtained with 14.4 keV  $^{57}\rm{Fe}$  solar axions \cite{Der11}) and is   stronger than our previous result obtained in $^{83}$Kr experiment \cite{Gav15}. As in the case
of $^{57}\rm{Fe}$ nucleus the obtained limit on axion mass strongly depends on the exact values of the parameters $S$ and $z$.

\section{Acknowledgments}

This work was supported by the Russian Foundation of Basic Research (grants 17-02-00305A, 16-29-13014ofi-m, 16-29-13011ofi-m, 15-02-02117A, 14-02-00258A).

% ****************************************************************************
% BIBLIOGRAPHY AREA
% ****************************************************************************

\begin{footnotesize}

\end{footnotesize}

% ****************************************************************************
% END OF BIBLIOGRAPHY AREA
% ****************************************************************************

\end{document}